\documentclass{DISproc}

\begin{document}

\title{Progress in the dynamical parton distributions}

\author{{\slshape Pedro Jimenez-Delgado}\\[1ex]
Institut f\"ur Theoretische Physik, Universit\"at Z\"urich, 8057 Z\"urich, Switzerland\\
Jefferson Lab, 12000 Jefferson Avenue, Newport News, VA 23606, USA}

\contribID{299}

\doi

\maketitle

\begin{abstract}
The present status of the (JR) dynamical parton distribution functions is reported. Different theoretical improvements, including the determination of the strange sea input distribution, the treatment of correlated errors and the inclusion of alternative data sets, are discussed. Highlights in the ongoing developments as well as (very) preliminary results in the determination of the strong coupling constant are presented.
\end{abstract}

The \emph{dynamical} parton distributions of the nucleon at $Q^2\!\gtrsim\!1$~GeV$^2$ are QCD radiatively generated from {\em valencelike}\footnote{Valencelike refers to $a_f\!>\!0$ for {\em all} input distributions $xf(x,Q_0^2)\propto x^{a_f}(1-x)^{b_f}$, i.e., not only the valence but also the sea and gluon input densities vanish at small $x$.} positive definite input distributions at an optimally determined low input scale $Q_0^2\!<\!1$~GeV$^2$. Thus the \emph{steep} small-Bjorken-$x$ behavior of structure functions, and consequently of the gluon and sea distributions, appears within the dynamical (radiative) approach mainly as a consequence of QCD-dynamics at $x \lesssim 10^{-2}$ \cite{Gluck:1994uf}. Alternatively, in the common ``standard'' approach the input scale is fixed at some arbitrarily chosen $Q^2_0\!>\!1$~GeV$^2$, and the corresponding input distributions are less restricted; for example, the mentioned {\em steep} small-$x$ behavior has to be {\em fitted}.

Following the radiative approach, the well-known LO/NLO GRV98 dynamical parton distribution functions of \cite{Gluck:1998xa} have been updated in \cite{Gluck:2007ck}, and the analysis extended to the NNLO of perturbative QCD in \cite{JimenezDelgado:2008hf}. In addition, in \cite{Gluck:2007ck, JimenezDelgado:2008hf} a series of ``standard'' fits were produced in (for the rest) exactly the same conditions as their dynamical counterparts. This allows us to compare the features of both approaches and to test the the dependence in model assumptions. The associated uncertainties encountered in the determination of the parton distributions turn out, as expected, to be larger in the ``standard'' case, particularly in the small-$x$ region, than in the more restricted dynamical radiative approach where, moreover, the ``evolution distance'' (starting at $Q_0^2\!<\!1$~GeV$^2$) is sizably larger \cite{Gluck:2007ck, JimenezDelgado:2008hf}.

The strong coupling constant $\alpha_s(M_Z^2)$ was determined in our analyses together with the parton distributions, in particular it is closely related to the gluon distribution which drives the QCD evolution and consequently its uncertainty is also smaller in the dynamical case. We obtained $\alpha_s(M_Z^2) =$ 0.1124 $\pm$ 0.0020 at NNLO, and 0.1145 $\pm$ 0.0018 at NLO in the dynamical case; to be compared with $\alpha_s(M_Z^2) =$ 0.1158 $\pm$ 0.0035 at NNLO, and 0.1178 $\pm$ 0.0021 at NLO in the ``standard'' one. The difference between these values is to be interpreted as a genuine uncertainty stemming from parameterization dependence, e.g. our contribution to the next PDG determination will be an average over NNLO dynamical and standard results. We consider this approach to be more realistic than considering only the reduced errors stemming exclusively from experimental uncertainties in each single analysis \cite{qsrole}.

The dynamical predictions for $F_L(x,Q^2)$ become perturbatively stable already at $Q^2\!=\!2$ - 3~GeV$^2$, where precision measurements could even delineate NNLO
effects in the very small-$x$ region. Moreover they are positive and in excellent agreement with the latest  H1 data \cite{Collaboration:2010ry}. This is in contrast to the results based in the common ``standard'' approach, which are less precise and in some cases turn negative at the lower $Q^2$ values.

The inclusive production cross sections for $W^+, W^-$ and $Z^0$-bosons form important benchmarks for the physics at hadron colliders. Our NLO \cite{Gluck:2008gs} and NNLO \cite{JimenezDelgado:2009tv} predictions for these standard candles processes were presented not long ago, and a detailed comparison of the predictions predictions based on the available NNLO parton parameterizations was performed in \cite{Alekhin:2010dd}. According to these studies the rates for gauge boson  production at the LHC can be predicted with an accuracy of better than about 10\% at NNLO. NNLO predictions for the Higgs boson production cross sections for Tevatron and LHC energies were also considered; the production rates could be predicted with an accuracy of about 10--20\% at the LHC. The inclusion of the NNLO contributions is mandatory for achieving such accuracies since the total uncertainties are substantially larger at NLO.

During the time passed since the last determination of our PDFs there have been experimental and theoretical developments which deserve consideration. From the experimental side, besides results coming from LHC which we will not discuss here, the combined HERA data on neutral current and charge current DIS has been published \cite{Aaron:2009aa}; the combination of the charm production data is ready but unfortunately still not publicly﻿ available. For the older fixed-target data in our analyses we use now directly the measured cross-sections (instead of the extracted structure functions) since potential problems with some of these extractions have been pointed out \cite{Alekhin:2011ey}. A wealth of deuteron data have also been included, for which description we use the nuclear corrections of \cite{Accardi:2011fa}. Further theoretical improvements include the use of the running mass definition for DIS charm and bottom production \cite{Alekhin:2010sv}.

In this proceeding we will concentrate in several concrete aspects of the ongoing update, modifications in the ansatz for the strangeness input distributions, and some consequences of the treatment of correlated systematic errors, in particular of the normalization uncertainties. Some preliminary results on determinations of  $\alpha_s(M_Z^2)$  and the (future) inclusion of higher-twist terms in our analyses are also briefly discussed.

\begin{figure}[htb]
  \centering
  \includegraphics[width=0.72\textwidth]{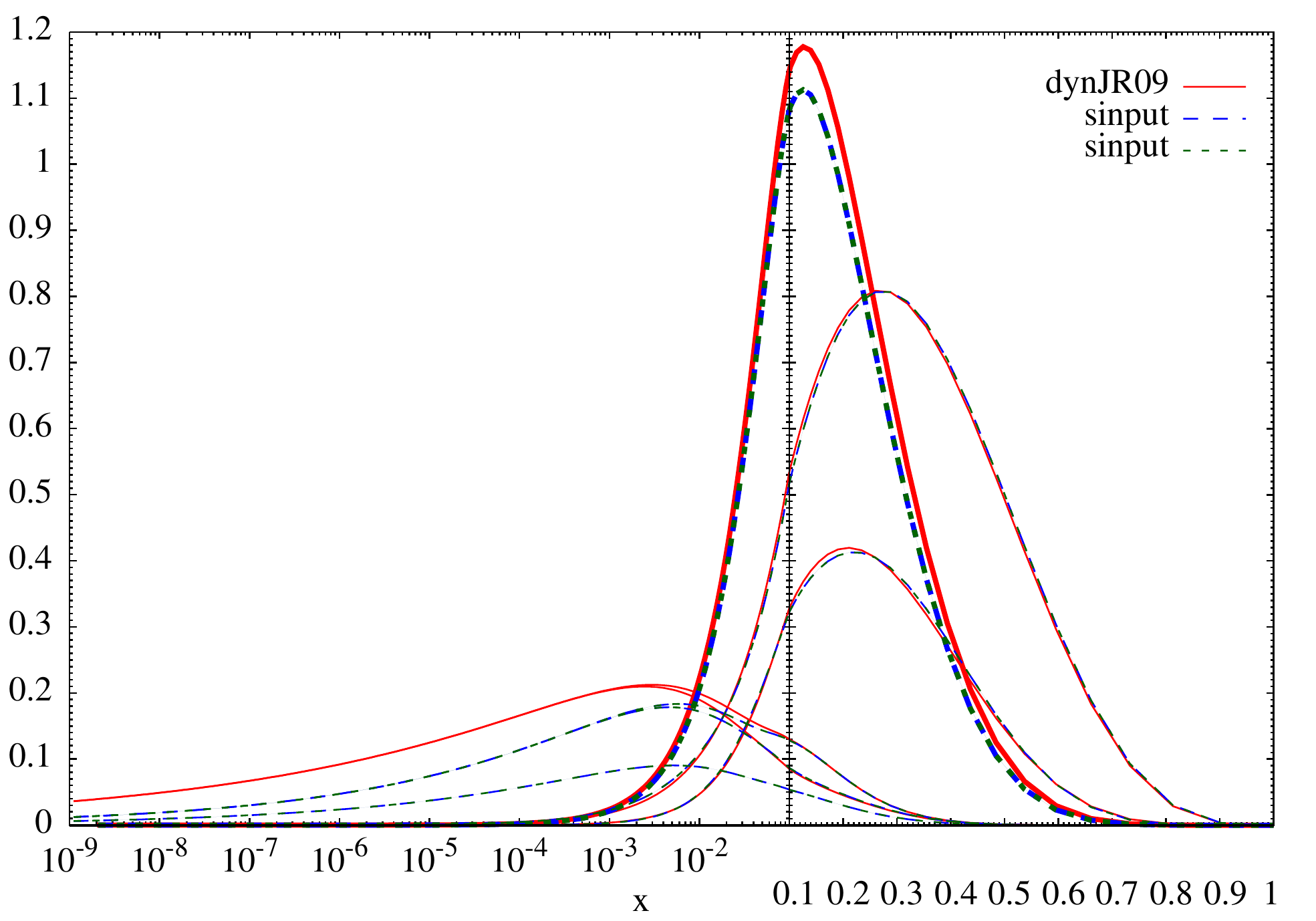}
  \caption{Dynamical NNLO input distributions (``sinput'') at $Q^2_0 = 0.55$ GeV$^2$ obtained using $s(x,Q_0^2)=\bar{s}(x,Q_0^2)= \tfrac{1}{4}\left(\bar{u}(x,Q_0^2)+\bar{d}(x,Q_0^2)\right)$. The original JR09 \cite{JimenezDelgado:2008hf} were obtained using $s(x,Q_0^2)=\bar{s}(x,Q_0^2)=0$ and are also shown for comparison.}
  \label{sinput}
\end{figure}

As in previous dynamical determinations \cite{Gluck:1994uf, Gluck:1998xa}, since the data sets used are insensitive to the specific choice of the strange quark distributions, the strange densities of the \emph{dynamical} distributions in \cite{Gluck:2007ck, JimenezDelgado:2008hf} have been generated entirely radiatively starting from vanishing strange input distributions:
\begin{equation}
s(x,Q_0^2)=\bar{s}(x,Q_0^2)=0
\label{nullstrange}
\end{equation}
at the \emph{low} input scale\footnote{In the ``standard'' case, where $Q^2_0\!>\!1$ GeV$^2$, the strange input distributions were chosen $s(x,Q_0^2)=\bar{s}(x,Q_0^2)= \tfrac{1}{4}\left(\bar{u}(x,Q_0^2)+\bar{d}(x,Q_0^2)\right)$, as is conventional \cite{Gluck:2007ck, JimenezDelgado:2008hf}.}. The plausibility of these \emph{assumptions} was investigated in \cite{JimenezDelgado:2010pc} by confronting the predictions derived from dynamical distributions determined in this way with the most precise dimuon production data \cite{Mason:2007zz}, which were the most sensitive to the strangeness content of the nucleon until the recent ATLAS measurements \cite{Aad:2012sb}. A good agreement was found at NLO. Since the NNLO corrections (thus the experimental acceptance corrections) for these process are not known, a consistent study of these data at NNLO is not possible. However it is apparent that unless the NNLO corrections were very large (which seems unlikely) our JR09 NNLO distributions would generally undershoot
the data. Indeed the would-be NNLO results favor $\bar{s}\simeq 0.1\left(\bar{u}+\bar{d}\right)$ at the input scale. In order to estimate the implications which this could have for our predictions for benchmark processes at the LHC we have repeated our JR09 dynamical analysis using  $s(x,Q_0^2)=\bar{s}(x,Q_0^2)= \tfrac{1}{4}\left(\bar{u}(x,Q_0^2)+\bar{d}(x,Q_0^2)\right)$ as in our ``standard'' fits. The resulting input distributions are compared with the original JR09 input in Fig.~\ref{sinput}. It can be observed that the increase in the strange distributions is \emph{compensated} by a decrease in the light sea, while the other distributions remain practically unaltered. The mentioned benchmark cross-sections for $W$ and $Z$ production are practically identical in both cases. Nevertheless the determinations of strange input distributions will have to be revisited in light of the mentioned ATLAS measurements \cite{Aad:2012sb}, which seem to imply a greater strange content of the nucleon.

Another aspect of our analyses in which we have been working is a more complete treatment of the correlated systematic uncertainties of the data. The least-squares estimator that we use to take them into account has been explicitly written down in Appendix B of \cite{Stump:2001gu}. This is used as well for normalization uncertainties, and implies that data and theory are relatively shifted by amounts which are determined by the minimization but not restricted further. This is in contrast to the treatment that we used in our (G)JR analyses, where the normalization shifts were limited by the experimental normalization uncertainties, and has resulted in small shifts on the valence distributions in the medium $x$ region.

Yet another issue raised by the ABM collaboration is the necessity of including higher twist for the description of fixed target data \cite{Alekhin:2012ig}; even if moderate kinematic cuts are used to select the data included in the fits; in particular for SLAC and NMC data. The kinematic cuts in our G(JR) analyses were $Q^2 \geq 4$ GeV$^2$ and $W^2\geq 10$ GeV$^2$, and were applied to the $F_2$ values extracted from different beam energies and combined. The description was good for NMC data and rather poor for SLAC data. However since the number of points for these experiments was rather small (about 100 data points for NMC and 50 for SLAC), the values of the cuts did not affect very much the results of the fits.

This picture changes if, as the ABM collaboration does, data on the cross sections for individual energies are used, which amounts to hundreds of data points for each experiment. As mentioned in the introduction, we also use these data in our current preliminary analyses, for which the virtuality cut has been raised to $Q^2 \geq 9$ GeV$^2$. We find results which are rather similar to our JR09 analyses, with $\alpha_s(M_Z^2)$ values about 0.113 to 0.114, depending on the input scale. If the usual cut of $Q^2 \geq 4$ GeV$^2$ is applied to these data $\alpha_s(M_Z^2)$ raises to 0.1176; a tendency in agreement with the ABM observations \cite{Alekhin:2012ig}. The inclusion of higher twist terms in the theoretical description should strongly reduce the dependence of the outcome of the fits and is currently under consideration.

To summarize, we have discussed several aspects of the ongoing update of the JR dynamical parton distributions, in particular the determination of strangeness input distributions, the treatment of normalizations, and results on $\alpha_s(M_Z^2)$ and their dependence on kinematic cuts and the treatment of higher-twist contributions. Changes in the parton distributions induced by all this modifications in the analysis and in the data have been, until now, quite modest.

{\raggedright\section*{Acknowledgements}
We thank E.~Reya for a fruitful collaboration, and S.~Alekhin, J.~Bl\"umlein and S.~Moch for discussions. This research is supported in part by the Swiss National Science Foundation (SNF) under contract 200020-138206. Authored by Jefferson Science Associates, LLC under U.S. DOE Contract No. DE-AC05-06OR23177. The U.S. Government retains a non-exclusive, paid-up, irrevocable, world-wide license to publish or reproduce this manuscript for U.S. Government purposes.

\begin{footnotesize}

\end{footnotesize}
}
\end{document}